\documentclass[twocolumn,showpacs]{revtex4}  
 \usepackage[latin1]{inputenc}
 \usepackage{amsmath}
 \usepackage{epsfig}
 \usepackage{subfigure}
 \usepackage{placeins}
 \usepackage{floatflt}
 \usepackage{amsmath}
 \usepackage{amssymb}
 \usepackage{upgreek}
 \usepackage{natbib}
 \usepackage{wrapfig}
 \usepackage[]{tocbibind}
 \usepackage{lineno}
 \usepackage{graphicx}
\usepackage{pst-all}

 \usepackage{ifthen}
 \usepackage{times}

\makeatletter
\newcommand\figcaption{\def\@captype{figure}\caption}
\makeatother

\newlength{\mylinewidth}
\begin{document}

\title{Baryonic resonances close to the ${\bf\bar{K}}$N threshold: the case of ${\bf\Sigma(1385)^+}$ in pp collisions}
\author{\footnotetext[2]{deceased}\footnotetext[1]{corresponding authors:}G.~Agakishiev$^{6}$, A.~Balanda$^{3}$,
D.~Belver$^{17}$, A.V.~Belyaev$^{6}$, A.~Blanco$^{2}$, M.~B\"{o}hmer$^{9}$, J.~L.~Boyard$^{15}$, P.~Cabanelas$^{17}$, E.~Castro$^{17}$, J.C.~Chen $^{8}$
% ...
}

\affiliation{
\footnotesize
\begin{center}
(HADES collaboration)\\
\end{center}
\\\mbox{$^{1}$Istituto Nazionale di Fisica Nucleare - Laboratori Nazionali del Sud, 95125~Catania, Italy}\\
\mbox{$^{2}$LIP-Laborat\'{o}rio de Instrumenta\c{c}\~{a}o e F\'{\i}sica Experimental de Part\'{\i}culas , 3004-516~Coimbra, Portugal}\\
\mbox{$^{3}$Smoluchowski Institute of Physics, Jagiellonian University of Cracow, 30-059~Krak\'{o}w, Poland}\\
\mbox{$^{4}$GSI Helmholtzzentrum f\"{u}r Schwerionenforschung GmbH, 64291~Darmstadt, Germany}\\
\mbox{$^{5}$Institut f\"{u}r Strahlenphysik, Helmholtz-Zentrum Dresden-Rossendorf, 01314~Dresden, Germany}\\
\mbox{$^{6}$Joint Institute of Nuclear Research, 141980~Dubna, Russia}\\
\mbox{$^{7}$Institut f\"{u}r Kernphysik, Johann Wolfgang Goethe-Universit\"{a}t, 60438 ~Frankfurt, Germany}\\
\mbox{$^{8}$ Excellence Cluster Universe, Technische Universit\"{a}t M\"{u}nchen, Boltzmannstr.2, D-85748, Garching, Germany}\\
\mbox{$^{9}$Physik Department E12, Technische Universit\"{a}t M\"{u}nchen, 85748~M\"{u}nchen, Germany}\\
\mbox{$^{10}$II.Physikalisches Institut, Justus Liebig Universit\"{a}t Giessen, 35392~Giessen, Germany}\\
\mbox{$^{11}$Istituto Nazionale di Fisica Nucleare, Sezione di Milano, 20133~Milano, Italy}\\
\mbox{$^{12}$Institute for Nuclear Research, Russian Academy of Science, 117312~Moscow, Russia}\\
\mbox{$^{13}$Frederick University, 1036 Nicosia, Cyprus}\\
\mbox{$^{14}$Department of Physics, University of Cyprus, 1678 Nicosia, Cyprus}\\
\mbox{$^{15}$Institut de Physique Nucl\'{e}aire (UMR 8608), CNRS/IN2P3 - Universit\'{e} Paris Sud, F-91406~Orsay Cedex, France}\\
\mbox{$^{16}$Nuclear Physics Institute, Academy of Sciences of Czech Republic, 25068~Rez, Czech Republic}\\
\mbox{$^{17}$Departamento de F\'{\i}sica de Part\'{\i}culas, Univ. de Santiago de Compostela, 15706~Santiago de Compostela, Spain}\\
\\
\mbox{$^{a}$ now at Lawrence Berkeley National Laboratory, Berkeley, USA}\\
\mbox{$^{b}$ also at ISEC Coimbra, Coimbra, Portugal}\\
\mbox{$^{c}$ also at ExtreMe Matter Institute EMMI, 64291 Darmstadt, Germany}\\
\mbox{$^{d}$ also at Technische Univesit\"{a}t Darmstadt, Darmstadt, Germany}\\
\mbox{$^{e}$ also at Technische Universit\"{a}t Dresden, 01062~Dresden, Germany}\\
\mbox{$^{f}$ also at Dipartimento di Fisica, Universit\`{a} di Milano, 20133~Milano, Italy}\\
\mbox{$^{g}$ also at Dipartimento di Fisica Generale and INFN, Universit\`{a} di Torino, 10125 Torino, Italy}}
\date{\today}
\begin{abstract}
We present results of an exclusive measurement of the first excited state of the $\Sigma$ hyperon, $\Sigma(1385)^+$, produced in $p+p \rightarrow \Sigma^+ + K^+ + n$ at $ 3.5\,\mathrm{GeV}$ beam energy.
The extracted data allow to study in detail the invariant mass distribution of the $\Sigma(1385)^+$.
The mass distribution is well described by a relativistic Breit-Wigner function with a maximum at $m_0=\,1383.2 \pm 0.9\, \mathrm{MeV/}c^2$ and a width of $40.2 \pm 2.1\, \mathrm{MeV/}c^2$. The exclusive production cross-section comes out to be $22.27 \pm0.89 \pm 1.56  ^{+3.07}_{-2.10}~\mu b$. Angular distributions of the $\Sigma(1385)^+$ in different 
reference frames are found to be compatible with the hypothesis that $33\,\%$ of
 $\Sigma(1385)^+$ result from the decay of an intermediate $\Delta^{++}$ resonance.
 \end{abstract}
\maketitle
\section{Introduction}
The excitation spectrum of hadrons reflects properties of quantum chromo dynamics (QCD) in the non-perturbative sector. Many hadrons, mesons as well as baryons, can be identified as members of multiplets within quark models. More advanced techniques, such as various effective theories which are directly related to QCD, however, generate dynamically certain resonances as emerging from interactions among other (more fundamental) hadrons \cite{eff}. This dual picture motivates the quest for understanding the very nature of resonances or hadrons in general.

Considering the $S=-1$ strange baryon sector, the $\Sigma(1385)$ is the first excited state of the $\Sigma$ hyperon and has a spin of $3/2\,\hbar$ and isospin $1$, in analogy to the $\Delta(1232)$ as first excited state of the nucleon. 
This resonance is characterized by a short life time that translates into a natural width of  $35.8 \pm\,0.8\,\mathrm{MeV/c}^2$ \cite{pdg}.
The $\Sigma(1385)$ itself is considered as a standard quark triplet but its vicinity to the $\Lambda(1405)$ in the mass spectrum correlates the study and understanding of the two resonances. 
Indeed, as pointed out in \cite{Lut08}, within a chiral SU(3) lagrangian approach the anti-kaon spectral distribution is intimately related to the $\Lambda$,  $\Lambda(1405)$, $\Lambda(1520)$, $\Sigma$ and $\Sigma(1385)$.
In particular, the $\Lambda(1405)$ arises naturally from hadronic effective models as a mixture of a $K^-N$ and $\pi\Sigma$ bound systems \cite{Jid03, Bor06, Hyo08} but the contribution of the two 'molecular' states and the dependency of the $\Lambda(1405)$ formation upon the reaction is still far from being understood.
Any analysis of the lowest lying  hyperon resonances suffers from the overlapping mass distributions, from possible interference effects through their $\bar{K}-N$ coupling, and from the common decay into $\Sigma-\pi$. In a first step of such an analysis we will determine the  $\Sigma(1385)^+$ production characteristics in the $\Lambda-\pi$ decay channel (this paper) and address $\Lambda(1405)$ production in a forthcoming publication.

The largest part of the present knowledge on the $\Sigma(1385)$ hyperon has been gained by employing photo-production on a proton target \cite{Kel11} (for the corresponding theoretical analysis cf. \cite{Oh08}), $K^-$p collisions \cite{Bau84,Aug81}, and pp collisions \cite{Kle70}. The latter measurement was accomplished at a beam momentum of $6\, \mathrm{GeV/c}$ corresponding to an excess energy, defined as the energy above the production threshold, of $\epsilon=\,830\,\mathrm{MeV}$.
The HADES collaboration has recently measured pp collisions at $3.5$ GeV beam energy. The $\Sigma(1385)^+$ is identified in the $K^+ n p \pi^- \pi^+$ final state via the $\Lambda \pi^+$ decay of the $\Sigma(1385)^+$ (BR= $88\,\%$), where the $\Lambda$ is identified by the $p\pi^-$ decay (BR= $63.9\, \%$). 
The present paper reports on this measurement. 

Thinking of the exclusive $\Sigma(1385)^+K^+n$ production in pp collisions the one-boson exchange model can be considered. In this framework several diagrams are conceivable: (i) $\pi^+$, $\rho^+$ exchange and $\Sigma(1385)^+$ production in a $\pi^+(\rho^+)p-\Sigma(1385)^+K^+$ vertex with possible excitation of an intermediate $\Delta^{++}$ decaying into $\Sigma(1385)^+K^+$; (ii) $K^{0(*)}$ exchange and production in the $pK^{0(*)}-\Sigma(1385)^+$ vertex. Another possibility is related to internal meson conversion in a $\pi^+K^+K^{0*}$ vertex and $\Sigma(1385)^+$ production in the $K^{0*}p-\Sigma(1385)^+$ vertex. Pion exchange diagrams have been considered in the analysis \cite{Kle70}, while in \cite{Ferr} kaon exchange has been included as well. Particularly interesting is the role of the intermediate $\Delta^{++}$ excitation, as discussed in \cite{Chi68}.
The angular distribution of the $\Sigma(1385)^+$ extracted in this work provides information which can serve to disentangle the different production mechanisms.

The results presented here have also to been seen in the context of a forthcoming analysis of the p(3.5 GeV)+Nb reaction, investigated with the same apparatus, which allows for studies of the medium modifications of the $\Sigma(1385)$. 
In the nuclear medium, the $\Sigma(1385)$ is predicted to suffer an attractive interaction encoded in the real part of the self-energy resulting in a $40\, \mathrm{MeV}$ "mass shift" and a broadening of the width by a factor of two at normal nuclear matter density \cite{Kas05}. 
For further discussions of in-medium modifications of $\Sigma(1385)$ see \cite{Cas03,Scha00,Lut04}. 
A first identification of the $\Sigma(1385)$ in sub-threshold heavy ion collisions has been reported in \cite{Lop07}, in which the statistical error of the extracted signal did not allow to draw conclusions about a possible broadening of the resonance spectral function. The measured yield of the $\Sigma(1385)$ \cite{Lop07} is well reproduced by a statistical model.

Baryon resonances, such as the $\Lambda(1520 )$  and the $\Sigma(1385)$, have also triggered the interest in the context of heavy ion reactions at higher energies \cite{Abe08,Mar08}. There, the measurement of the resonance abundance is thought to reflect the dynamics of their production and can be connected with a stage of the reaction prior to hadronisation \cite{Vog10}.

Following this line of reasoning, we present here an analysis of the spectral shape of the $\Sigma(1385)^+$ reconstructed in the exclusive reaction $p(3.5 \,\mathrm{GeV})+p\rightarrow \Sigma(1385)^+ + K^++n$. 
Our work is organized as follows. Sections II and III present the experimental set-up together with the event and particle selection. Section IV deals with techniques developed to identify the different background sources, and in section V the extracted $\Sigma(1385)^+$ signal is discussed. Sections VI and VII deal with the differential cross section and the modeling of the acceptance corrections, respectively. We close with a summary and conclusions in section VIII.

\section{The Experiment}
The experiment was performed with the
{\bf H}igh {\bf A}cceptance {\bf D}i-{\bf E}lectron {\bf S}pectrometer (HADES)
at the heavy-ion synchrotron SIS18
at GSI Helmholtzzentrum f\"ur Schwerionenforschung
in Darmstadt, Germany. HADES is a charged-particle detector consisting of a 6-coil toroidal magnet centered on the beam axis and six
identical detection sections (with nearly complete azimuthal coverage) located between the coils and covering polar angles between 18$^{\circ}$ and 85$^{\circ}$. Each sector is
equipped with a Ring-Imaging Cherenkov (RICH) detector followed by Multi-wire Drift Chambers (MDCs), two
in front of and two behind the magnetic field, as well as two scintillator hodoscopes (TOF and TOFino). Lepton identification
is provided mostly by the RICH and supplemented at low polar angles with Pre-Shower detectors, mounted
at the back of the apparatus. Hadron identification is based on the time-of-flight and on energy-loss
information from the scintillators and the MDC tracking detectors. In the following the TOF-TOFino-PreShower system is referred to as Multiplicity And Time-of-flight Array (META). A detailed description of the spectrometer can be found in
\cite{hadesSpectro}. 

A proton beam of $\sim 10^7$ particles/s with $3.5\, \mathrm{GeV}$ kinetic energy was incident on a liquid hydrogen target of $50\, \mathrm{ mm}$ thickness corresponding to $0.7\,\%$ interaction length.  
The data readout was started by a first-level trigger (LVL1) requiring a charged-particle multiplicity, $\mathrm{MUL}\,>3$, in the META system.
A total of $1.14\times10^9$ events was recorded under these experimental conditions.

A common start time for all particles measured in one event was reconstructed as described in detail in \cite{Sieb}. 
With this method the time-of-flight information could be determined for almost all of the measured events.
\section{Particle Identification and Event Selection}
\label{id1}
The aim of this analysis is to reconstruct the $\Sigma(1385)^+$ signal in the following reaction and decay chain:
\begin{flushleft} 
\label{reac1}
\begin{flalign} 
\begin{pspicture}(0.,0.)
  \psline[ArrowInside=-]{->}(3.5,-0.1)(3.5,-0.45)(3.9,-0.45)
    \psline[ArrowInside=-]{->}(4.1,-0.7)(4.1,-1.05)(4.5,-1.05)
     \end{pspicture}
 p(3.5GeV)+p  \rightarrow \Sigma(1385)&^+ + K^{+}+n   \label{S1385PProduction}\\
                                        &\Lambda+\pi^{+}                 \nonumber  \\   
                                        &\qquad p+\pi^{-}.  \hspace{0.5cm}   \nonumber                                                                         \end{flalign}
\end{flushleft}
The value of BR= $\approx\,56\, \%$ accounts for the total branching ratio via this decay chain.

\begin{figure}[!htb]
\begin{center}
\includegraphics[viewport= 10 3 336 500,angle=0,scale=0.48]{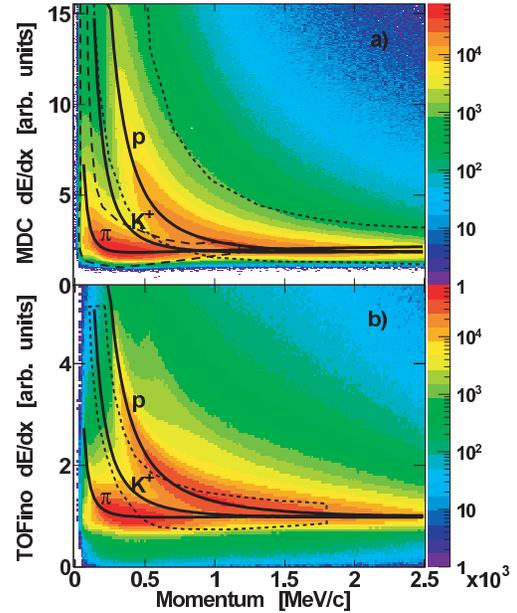}
\caption[]{Color online. $dE/dx$ versus momentum for all detected particles in the MDC (panel (a)) and TOFino (panel (b)) detectors. The solid black curves indicate adapted Bethe-Bloch formulas and the dashed ones show the interval used to select protons and pions in the MDC (panel (a)) and kaons in TOFino (panel (b)).}
\label{dedx}
\end{center}
\end{figure}

The first analysis step consists of selecting events containing four charged particles ($p$, $\pi^-$, $\pi^+$, K$^+$).
Particle identification is performed employing the energy loss ($dE/dx$) of protons and pions  in the MDCs and adding the $dE/dx$ information from the TOFino system to further select kaons. The kaon selection is performed by TOFino only,  because only a negligible amount of kaons are observed in the TOF region. Fig.~\ref{dedx} shows the $dE/dx$ distribution as a function of the particle momentum extracted from the MDCs (panel (a)) and TOFino (panel (b)). The full black curves refer to adapted Bethe-Bloch formula \cite{PhD_Schmah} and the dotted and dashed lines show the cuts applied to select pion and proton candidates in the MDC: (panel (a)) and kaons in the TOFino (panel (b)).
The masses of these kaon candidates are calculated using momentum and velocity. The purity of the kaon signal is then enhanced by
selecting masses between $280-780\, \mathrm{MeV/}c^{2}$.
The reconstruction of the $\Lambda$ hyperon is achieved exploiting the decay $\Lambda\rightarrow p+\pi^-$ (BR= $63.9\, \%$). Since the $\Lambda$ hyperon has a mean decay length of c$\tau\approx$ 7.89 cm, selections on the decay vertex can be applied to reduce the background.

\begin{figure}[!htb]
\begin{center}
\includegraphics[viewport= 30 13 536 420,angle=0,scale=0.42]{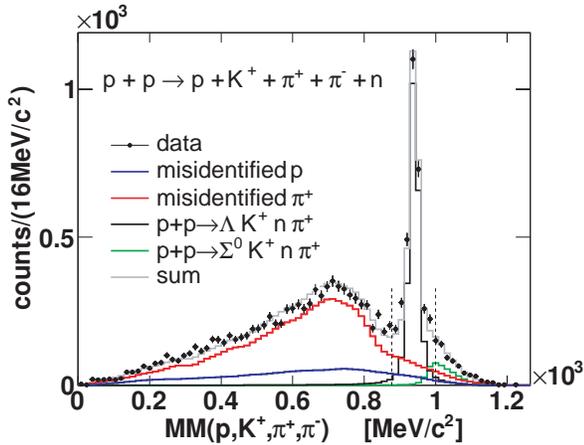}
\caption[]{Color online. Missing mass to the four charged particles p, $\pi^-$, $\pi^+$ and K$^+$, after the selection of a $\Lambda$ candidate in the same event. The gray full curve shows the sum of all identified contributions.}
\label{neuMM}
\end{center}
\end{figure}

The primary reaction vertex has been estimated by computing the point of closest approach between the $\pi^+$ and the K$^+$ track pairs.
The following topological cuts are employed to select $\Lambda$ candidates and to enhance the signal-to-background ratio in the $p$-$\pi$ invariant mass distribution: (1) distance between the proton and pion tracks ($d_{p-\pi{^-}}<\,20\, \mathrm{mm}$), (2) distance of closest approach to the primary vertex for the proton ($\mathrm{DCA_p}>\, 0 \, \mathrm{mm}$) and for the pion ($\mathrm{DCA_{\pi^-}}>\,\mathrm{DCA_p}$), (3) distance between the primary reaction vertex and the secondary decay vertex ($d(\Lambda -V)\geq\, 15\, \, \mathrm{mm}$).
\begin{figure}[!htb]
\begin{center}
\includegraphics[viewport= 30 33 536 420,angle=0,scale=0.42]{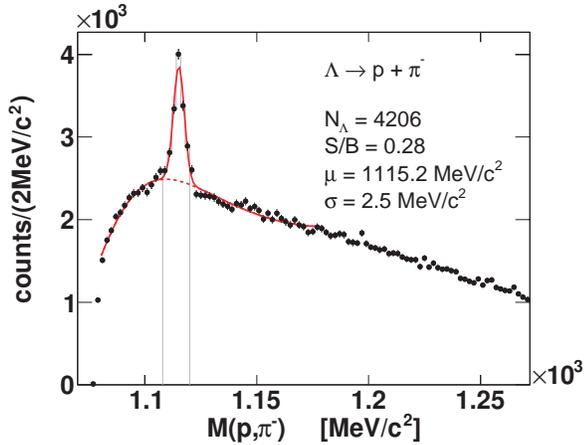}
\caption[]{Color online. Invariant mass distribution of selected $p$-$\pi^-$ pairs after the K$^+$ selection (see text for details). The two vertical lines define the mass interval select events with $\Lambda$ candidates.}
\label{lamMass}
\end{center}
\end{figure}
The resulting invariant-mass distribution of the selected p-$\pi^-$ pairs is shown in Fig.~\ref{lamMass}, in which the contribution of the $\Lambda$ hyperon is clearly visible. The background was fitted using a Landau function plus a polynomial of third order, while a Gaussian fit was applied to the signal. The obtained mean value and variance for the reconstructed mass are $M_{\Lambda}=\,1115.2\, \mathrm{MeV/}c^2$ and $\sigma_{\Lambda}=\,2.5\,\mathrm{MeV/}c^{2}$, respectively.
In order to further improve the data sample, only those events are selected for which the $p$-$\pi^-$ invariant mass is within $1110-1120 \,\mathrm{MeV/}c^2$ (corresponding to 2.0 $\sigma$), as indicated in Fig.~\ref{lamMass} by the vertical lines.
A further step consists in the selection of the missing neutron appearing in reaction (\ref{S1385PProduction}), which cannot be detected in HADES.
The missing mass to the four charged particles ($p$, $\pi^-$, $\pi^+$, K$^+$) is calculated after the $\Lambda$ and K$^+$ selection and the result is shown in Fig.~\ref{neuMM}.
A clear signal close to the nominal neutron mass is visible (M$_{n}= \,939.5\,\mathrm{MeV/}c^2$, $\sigma_{n}=\, 12.6\,\mathrm{MeV/}c^2$) sticking out of a broad background distribution.
Simulation studies have shown that the background to the left of the neutron peak is entirely caused by events with protons and pions misidentified as kaons. The red and the blue histograms in Fig.~\ref{neuMM} show the background due to pion and proton misidentification obtained by a side-band analysis of the experimental data, as discussed in section \ref{backS}.
 Additionally, Fig.~\ref{neuMM} illustrates the contribution to the neutron signal from the non-resonant channel $p+p\rightarrow \Lambda+K^++n+\pi^+$ (black histogram) and  $p+p\rightarrow \Sigma^0+K^++n+\pi^+$ (green histogram). 
While the shape of these two contributions is obtained by full-scale simulations, the absolute yield of these channels
can be determined by fitting the total experimental spectrum.

\begin{figure}[!htb]
\begin{center}
\includegraphics[viewport= 30 13 536 420,angle=0,scale=0.42]{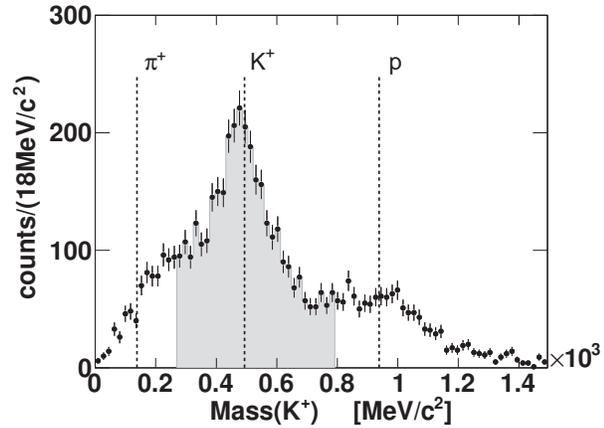}
\caption[]{Mass distribution of the selected K$^+$ after the cuts on the $\Lambda$ and missing neutron candidates (see text for details). The vertical lines represent the nominal masses of $\pi^+$, K$^+$ and proton respectively. The gray-shaded area shows the applied selection on the K$^+$ mass.}
\label{kaonM}
\end{center}
\end{figure}

Fig.~\ref{kaonM} shows the K$^+$ mass distribution after the $\Lambda$ selection and a cut around the neutron mass $877\, \mathrm{MeV/}c^2<\,MM(p,K^+,\pi^+,\pi^-)< \,999 \,\mathrm{MeV/}c^2$. This cut is shown by the vertical dashed lines in Fig.~\ref{neuMM} and corresponds to $\pm\,5\,\sigma$.  
\section{Background Identification}
\label{backS}
\begin{figure}[!htb]
\begin{center}
\includegraphics[viewport= 30 13 536 420,angle=0,scale=0.42]{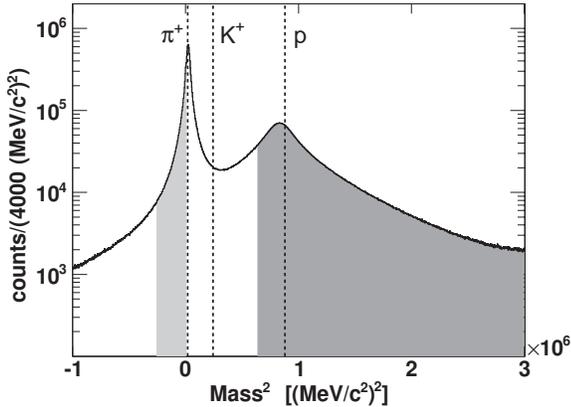}
\caption[]{Count rate of all K$^+$ candidates as a function of the squared mass. The two gray-shaded areas show the selected $\pi^+$ and the protons on the left and right side, respectively. The dashed verticals line indicate the nominal squared masses of $\pi^+$, K$^+$ and protons. }
\label{allK}
\end{center}
\end{figure}

Due to the incomplete purity of the kaon selection, even after all the cuts applied so far, a large fraction of the kaon candidates are still pions or protons. In particular, the contamination due to misidentified pions is rather high as visible in Fig.~\ref{kaonM}. In order to estimate quantitatively this background contribution a dedicated side-band analysis based on the mass distribution of the selected K$^+$ candidates has been developed. The goal of this analysis is to produce an event sample, which does not contain any kaon but either a pion or a proton misidentified as a kaon.
To emulate the background, events have been selected containing four charged particles: one proton, one $\pi^-$, one $\pi^+$ and a fourth positively charged particle for which no identification is required. The mass distribution of this fourth particle is shown in Fig.~\ref{allK}, where one can see that most of the yield is located around the nominal pion and proton masses and that no kaon peak is visible. 
These events represent a sample with a dominant contribution of misidentification background.
 A data sample with misidentified pions is explicitly extracted from the distribution in Fig.~\ref{allK} by selecting the mass range of $-0.25-0$ (GeV/c$^2$)$^2$, while for protons the interval from $0.615$ to $4\,(\mathrm{GeV/}c^2)^2$  is chosen. These background samples are analyzed in the same way as described in section~\ref{id1} with an exception made for the secondary vertex cuts of the $\Lambda$ decay. These cuts ((1-3) in section~\ref{id1}) have been omitted in order to enhance the statistics of the background sample. 

However, although the selection of the misidentified pions and protons is done in a range rather distant from the nominal kaon mass, the two misidentification samples still contain a certain amount of real kaons.
These real kaons are produced together with a $\Lambda$, due to strangeness conservation.
Since the selection on the $\pi^-$-proton pairs invariant mass around the $\Lambda$ nominal mass is also applied to the background samples, the contribution by real kaons is enhanced.

Our strategy to reduce the contribution of the real kaon signal to this background samples is to smear simultaneously the momenta of the identified $\pi^+$ and $\pi^-$ such that the $\Lambda$ signal in the $\pi^-$-proton invariant mass distribution disappears. This way, the$ \Lambda$ selection in the $p$-$\pi$ invariant mass distribution does not enhance the fraction of events with a $K^+$.

A further issue to be addressed deals with the specific kinematics of the particles selected via the cut around the nominal kaon mass (from here on considered as fake kaons) and of the explicitely misidentified pions and protons, employing the side band cut. The fake kaons correspond to particles with the 'correct' kaon mass but which are not real kaons. They are contributing to the missing mass distribution shown in Fig.~\ref{neuMM} mainly in the range away from the nominal neutron mass and are hence a source of background. 
\begin{figure}[!htb]
\begin{center}
\includegraphics[viewport= 30 13 536 420,angle=0,scale=0.42]{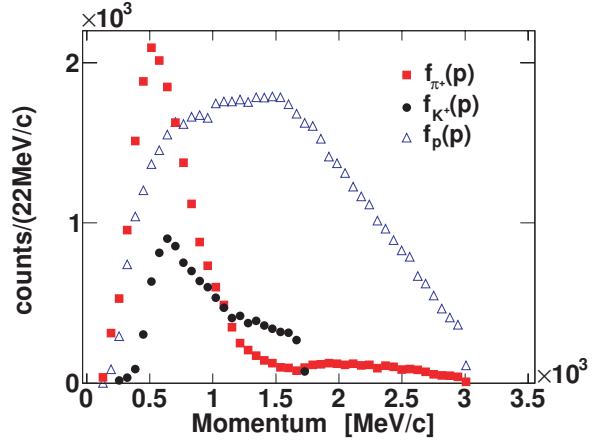}
\caption[]{Color online. Momentum distribution of the fake kaons (bullets) together with the protons (triangles) and pions (squares) explicitly misidentified as kaons.}
\label{momAll}
\end{center}
\end{figure}
Figure~\ref{momAll} compares the momentum distribution of the fake kaons (bullets) together with the momentum distributions of the misidentified protons (triangles) and pions (squares). One can see that the shapes of these distributions are rather different. This difference cannot be neglected during the evaluation of background due to fake kaons, since we are performing an exclusive analysis with the total four-momentum being conserved.\\
In order to reproduce the momentum distribution of the fake kaons the following function must be determined:
\begin{equation}
\gamma (|\vec{p}|) =\frac{f_{K^+}(|\vec{p}|)}{a_{\pi^+}f_{\pi^+}(|\vec{p}|)+f_{p}(|\vec{p}|)},
\end{equation}
where f$_{K^+}(|\vec{p}|)$, f$_{\pi^+}(|\vec{p}|)$  and  f$_{p}(|\vec{p}|)$ are the momentum distributions of the fake kaons, misidentified pions and protons, respectively. The factor $a_{\pi^+}$ accounts for the unknown relative contribution of the misidentified pions and protons to the total background. Given a value of the parameter $a_{\pi^+}$, each event corresponding to the pion 'side' of the mass distribution in Fig.~\ref{kaonM} is weighted by the factor $a_{\pi^+}$ and multiplied by $\gamma(|\vec{p}|)$, while each event from the proton 'side' is weighted with $\gamma(|\vec{p}|)$ only. 
The total missing mass is calculated for each weighted event and the obtained distributions are compared to the background underlying the neutron signal. Figure~\ref{neuMM} shows the resulting missing mass generated by the two background samples (blue histogram for protons and red histogram for pions). 
The parameter $a_{\pi^+}$ is varied systematically until the sum of the two background distributions gives the best fit to the region of the spectrum on the left side of the neutron peak, i. e. for $0\,< MM(p,K^+,\pi^+,\pi^-) <\,870\,\mathrm{MeV/}c^2$.
 Together with the such derived background  the contribution of the signal channels $p+p\rightarrow \Lambda+K^++n+\pi^+$ and $p+p\rightarrow \Sigma^0+K^++n+\pi^+$ are displayed in Fig.~\ref{neuMM} (black and green histograms respectively). The gray curve in Fig.~\ref{neuMM} shows the sum of all contributions, which reproduces the experimental data very well.
Our side-band method allows indeed to reproduce the total background. Furthermore, the contamination by misidentified pions and protons to the signal can be evaluated quantitatively. 
After the cut around the nominal neutron mass in the total missing mass distribution one obtains precisely the signal-to-background ratio and hence can estimate the contribution of the misidentification background to the $\Sigma(1385)^+$ spectrum. 
\section{The ${\bf\Sigma(1385)^+}$ Signal}
\label{sig}
\begin{figure}[!htb]
\begin{center}
\includegraphics[viewport= 30 13 536 420,angle=0,scale=0.42]{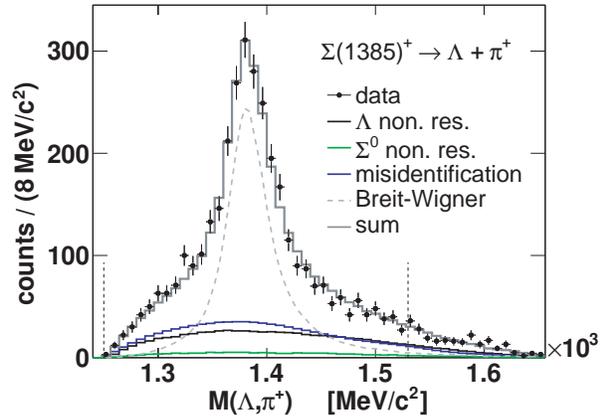}
\caption[]{Color online. Invariant mass distribution of the selected $\Lambda$-$\pi^+$ pairs. The circles show the experimental data, the black and green curves depict the contributions coming from the non-resonant production, the blue curve shows the misidentification background and the gray dashed curve represents a Breit-Wigner distribution fitted to the data. The solid gray histogram represents the result of the final fit (see text for details). Dashed vertical lines depict the interval chosen for the fit.}
\label{sigmaRaw}
\end{center}
\end{figure}

In the final step events conforming to the $\Lambda$, $\pi^+$, $K^+$ and neutron hypothesis are used to calculate the $\Lambda$-$\pi^+$ invariant mass.
The resulting distribution is shown in Fig.~\ref{sigmaRaw}. A clear peak at the $\Sigma(1385)^+$ pole mass is visible on the top of some background. 
The contribution of the misidentification background (blue histogram) is shown together with the contributions of the direct reactions (i) $p+p\rightarrow \Sigma^0+\pi^+ +K^++n$ (green histogram) and  (ii) $p+p\rightarrow \Lambda+\pi^+ +K^++n$ (black histogram).
After the determination of the scaling factors $a_{\pi^+}$ and $\gamma(|\vec{p}|)$ the form and yield of the misidentification background is fixed.
The reactions (i) and (ii) represent a source of background with the same hadrons in the final state as the reaction (1).  The absolute yield of the channel (i) is determined by fitting the signal in the neutron mass spectrum shown in Fig.~\ref{neuMM}, as already mentioned, while the yield of channel (ii) can only be determined by fitting the total distribution shown in Fig.~\ref{sigmaRaw}. 
We chose to fit the experimental $\Sigma(1385)^+$ signal with a Breit-Wigner function folded with the efficiency and acceptance corrections thus fitting the data instead of correcting the extracted experimental signal. This method allows us also to evaluate the contribution of the background source (ii) more precisely.
The used p-wave relativistic Breit-Wigner function \cite{Jack} reads as follows:
\begin{eqnarray}
\label{BWfun}
\text{Breit-Wigner} &\propto &\frac{q^2}{q^2_0}\frac{m^2_0\Gamma^2_0}{(m^2_0-m^2)^2+m^2_0\Gamma^2},  \\ \label{laura}
\Gamma&=&\Gamma_0 \frac{m_0q^3}{mq_0^3}F_1(q), \nonumber\\
 F_1(q)&=&\frac{1+(q_0R)^2}{1+(qR)^2},\nonumber
\end{eqnarray}
where $q$ is the momentum of the decay products in the $\Sigma(1385)^+$ rest frame, $q_0$ the momentum that corresponds to the pole mass $m_0$, $m$ the mass variable, $\Gamma_0$ the resonance width, $\Gamma$ the mass-dependent width, $F_1$(q) the Blatt-Weisskopf parameter and $R=\,1/197.327\,\mathrm{MeV}^{-1}$ the centrifugal barrier parameter. Accordingly, $q$, $q_0$, $m$, $m_0$, $\Gamma$ and $\Gamma_0$ are in MeV units here. 
The Blatt-Weisskopf centrifugal-barrier parameters \cite{Hip72} absorb possible divergences of the energy dependent width.
In order to quantify the different contributions to the spectrum in Fig.~\ref{sigmaRaw}, a fit is applied to the data in the mass range $1250\, \mathrm{MeV/}c^2<\,M(\Lambda-\pi^+)<\,1530\, \mathrm{MeV/}c^2$ (vertical dashed lines in Fig.~\ref{sigmaRaw}). The total fit function is composed of:
\begin{itemize}
\item the fixed distribution of the misidentification background and of channel (i),
\item the phase-space distribution of reaction (ii) filtered through the geometrical acceptance and the detector efficiency scaled by one fit parameter and
\item a corrected relativistic Breit-Wigner function with fit parameters for the mass, width and height.
\end{itemize}

The correction of the Breit-Wigner function is necessary to account for the finite available phase space and for the effects of the geometrical acceptance of the spectrometer and the efficiency of the analysis cuts that modify the measured spectral distribution in comparison to a pure Breit-Wigner function. The geometrical acceptance and reconstruction efficiencies for the $\Sigma(1385)^+$ have been estimated employing a Breit-Wigner distribution of the resonance as the input of a full-scale simulation. 
The procedure applied to extract the correct acceptance correction for the $\Sigma(1385)^+$ signal is explained in detail in sections~\ref{modAcce}.
These corrections also account for the finite resolution of the apparatus which cause a broadening of the $\Sigma(1385)^+$ resonance of about $3\, \mathrm{MeV}$. Since the fitted Breit-Wigner distribution is filtered through this correction, the extracted width and pole mass are already corrected also for the resolutions effects. 
In addition, the phase-space correction factor as a function of the invariant mass must be taken into account.
\begin{figure}[!htb]
\begin{center}
\includegraphics[viewport= 30 13 536 420,angle=0,scale=0.42]{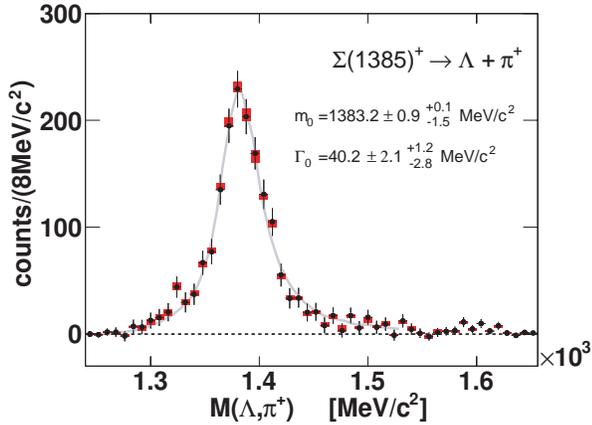}
\caption[]{Color online. Invariant mass distribution of the selected $\Lambda-\pi^+$ pairs after the background subtraction. The black circles show the experimental data together with the statistical and systematic (red squares) errors. The gray curve shows the fit with a p-wave Breit-Wigner function.}
\label{sigmaPure}
\end{center}
\end{figure}
This factor accounts for the loss of available phase space for the production of the neutron and K$^+$ in the reaction $p+p \rightarrow \Sigma(1385)^+ + K^+ + n$  with increasing  $\Lambda$-$\pi^+$ invariant mass and has been evaluated via simulations. The product of these two corrections is folded with the Breit-Wigner distribution \cite{Jack}. The product of this Breit-Wigner distribution with the correction factors is shown by the gray dashed curve in Fig.~\ref{sigmaRaw}.
The  final result of the fit is also shown in Fig.~\ref{sigmaRaw} by the gray histogram representing the sum of all contributions. It delivers a normalized $\chi^2$ value of about 1, a mass for the $\Sigma(1385)^+$ resonance of $1383.2 \pm 0.9\, \mathrm{MeV/}c^2$ and a width of $\Gamma_0= \,40.2 \pm 2.1\, \mathrm{MeV/}c^2$, where the errors are only statistical. Also other, commonly used, parametrisations of the Breit-Wigner have been tested and they have delivered results all compatible within statistical errors
The signal, after the subtraction of the three background contributions, is shown in Fig.~\ref{sigmaPure} together with the  relativistic p-wave Breit-Wigner function (\ref{BWfun}) obtained by fitting the distribution shown in Fig.~\ref{sigmaRaw}.
 The experimental data are displayed together with the statistical and systematic errors. The systematic errors have been evaluated modifying the analysis cuts as follows:
 the width of the cuts around the K$^+$, $\Lambda$ and neutron missing mass have been varied by $\pm 20$ \%. Additionally, the cuts applied to reconstruct the $\Lambda$ have been varied ($d_{p-\pi{^-}}<~ 28\, \mathrm{mm}$ and $d(\Lambda -V)\geq \,9\, \mathrm{mm}$, see section \ref{id1}).
For each set of cut values the same fitting procedure has been applied and a signal has been extracted after the background subtraction. The obtained systematic uncertainty of the signal is shown in Fig.~\ref{sigmaPure} by the red squares and lead to the following values for the pole mass and width of the resonance: $m_0=\,1383.2\pm 0.9 ^{+0.1}_{-1.5}\, \mathrm{MeV/}c^2$ and $\Gamma_0=\,40.2\pm2.1\,^{+1.2}_{-2.8}\, \mathrm{MeV/}c^2$. The value of the width is about $4\, \mathrm{MeV/}c^2$ larger than the PDG average quoted in \cite{pdg}.
On the other hand, the results obtained with this analysis take into account all the kinematic and efficiency effects that might modify the spectral function. The larger width of the $\Sigma(1385)^+$ resonance might also be due to its specific production mechanism in p+p collisions. Indeed, all the results collected in the PDG \cite{pdg} refer to $K^-p$ and $\pi ^+p$ reactions, where only a limited set of $K^-p$ data were used to arrive to the quoted average. 

Recently the CLAS collaboration published results on the $\Sigma(1385)\rightarrow \Lambda+\gamma$ \cite{Kel11} measured in photon-induced reactions; but no detailed analysis of the resonance mass distribution is reported upon. 

\section{Differential cross sections}
The determination of the (differential) production cross sections requires
extrapolation to full phase space. The corresponding (differential) acceptance corrections are obtained via a full scale simulation. Its input distributions are generated by a suitable model and tuned
(iteratively) such that the model reproduces satisfactorily the obtained differential cross sections. The differential cross sections have been calculated in three reference frames which are
described in the next section.\\
\subsection{Reference frames and angles}
The exclusive cross section in a $2\rightarrow 3$ reaction with unpolarized particles, $d\sigma /(\frac{d^3p_a}{E_a}\frac{d^3p_b}{E_b}\frac{d^3p_c}{E_c})$, depends on four independent variables at a given value of $\sqrt{s}$. In fact, for the reaction $1+2\rightarrow a+b+c$, the 9-dimensional exit phase space is constrained on a 5-dimensional hypersurface due to energy and momentum conservation; azimuthal symmetry around the beam axis reduces the dimension of the exit phase space to four. 
Several choices of these four variables in different reference frames are conceivable \cite{MSW11}:
\begin{itemize}
\item the opposite-momentum frame in the entrance channel ($\vec{p_1}=\,-\vec{p_2}$) coinciding here with the $1$-$2$ center-of-mass system,
\item an opposite momentum frame in the exit channel ($\vec{p_a}=\,-\vec{p_b}$).
\end{itemize}
In the first case, one may consider the angular distribution of the particle $c$ with respect to the beam axis, yielding a $\Theta_{1-2}^c$ distribution. In the second case, one may consider the angular distribution of particle $c$ with respect to the $a-b$ direction yielding a $\Theta^c_{a-b}$ distribution or with respect to particle 1 yielding a $\Theta^1_{a-b}$ distribution. $\Theta^c_{1-2}$ is usually labeled as CMS distribution of particle $c$, $\Theta^c_{a-b}$ refers to the helicity angle and $\Theta_{a-b}^1$ is the Gottfried-Jackson angle. The CMS and the Gottfried-Jackson angles connect entrance and exit channels, while the helicity angles quantify relations within the exit channel. As the selection of particle labels $a$, $b$, $c$ is arbitrary for different species, there are three different choices for each of the angles $\Theta_{1-2}^c$, $\Theta_{a-b}^1$ and $\Theta_{a-b}^c$. It has to be noticed that for identical particles 1 and 2, one should average $\Theta_{a-b}^1$ and $\Theta_{a-b}^2$. For different hadrons $a$ and $b$ we supplement the superscript by the label of the hadron relative to which the angle is measured, e.~g., $\Theta^{1b}_{a-b}$ or  $\Theta^{c-b}_{a-b}$ etc. 
 
The helicity angle distribution represents a special projection of a Dalitz plot. In particular a uniformly populated Dalitz plot results in an isotropic helicity angle distribution, whereas physical or kinematical effects distorting the Dalitz plot reflect themselves in anisotropic helicity angle distributions. 

The motivation for an analysis within a Gottfried-Jackson frame arises from considering e. g. the distribution of the angle $\Theta^{\Sigma^*-p}_{\Sigma^*-K^+}$, which is the angle of the $\Sigma(1385)^+$ relative to the proton in the $\Sigma(1385)^+$-$K^+$ reference frame. This angle can give insight into the scattering process especially concerning the involved partial waves. This statement holds true also if an intermediate resonance $\Delta^*$ is excited, e.g. $\pi p\rightarrow \Delta^*\rightarrow\Sigma(1385)^+K^+$. 
\subsection{Modeling of the acceptance correction}
\label{modAcce}
\paragraph{Angular distributions in the center-of-mass system.}

To extract the differential cross sections as a function of
the CMS angle, the data sample was divided into 7 angular bins and the analysis procedure described in sections \ref{id1} and \ref{backS} was applied. 
This means that for each bin the neutron missing mass spectrum was evaluated and the background was determined, as described in section \ref{backS}. After the selection of the neutron, the $\Sigma(1385)^+$ signal was extracted from the ($\Lambda$-$\pi^+$) invariant mass distribution. 
\begin{figure}[!htb]
\begin{center}
\includegraphics[viewport= 30 13 506 400,angle=0,scale=0.42]{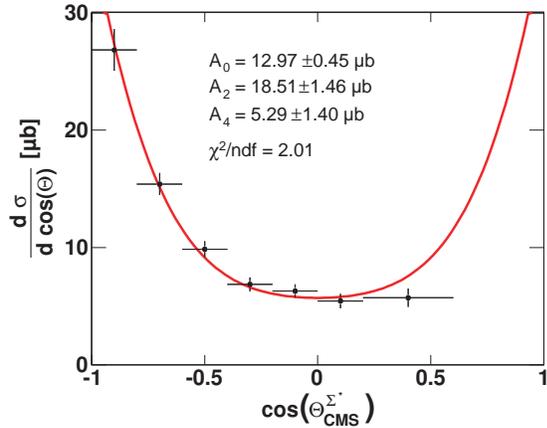}
\caption[]{Color online. Angular differential cross section for the $\Sigma(1385)^+$ production as a function of  $cos(\Theta^{\Sigma^* }_{CMS})$. The curve corresponds to a fit with a Legendre polynomials. }
\label{sigmaCM}
\end{center}
\end{figure}
The misidentification background was again fixed by the fit to the neutron spectra for each of the 7 angular bins. The background contributions due to the channels $p+p\rightarrow \Lambda+K^+ + n+\pi^+$ and $p+p\rightarrow \Sigma^0+K^++n+\pi^+$  were also fixed by using the scaling factor from the fit to the integrated spectrum (see Fig.~\ref{sigmaRaw}). In this way the background contribution could be subtracted, and a pure $\Sigma(1385)^+$ spectrum was obtained for each bin. Applying this binning also to the full-scale simulations, the geometrical acceptance and efficiency could be extracted in a differential form and could be used to correct the differential distributions. 

 The yield of $\Sigma(1385)^+$ in each angular bin was obtained by integrating the subtracted spectra in a $\pm\, 6\sigma$ interval around the pole mass, where the parameters were taken from the fit to the integrated spectrum (Fig.~\ref{sigmaPure}). 
 The production cross section was normalized to elastic scattering which was independently measured in the same experiment \cite{Ana10}.
Figure~\ref{sigmaCM} exhibits the differential cross section for $\Sigma(1385)^+$ production as a function of the CMS scattering angle.
The experimental data points shown in Fig.~\ref{sigmaCM} point to a strong anisotropy for the emission of the $\Sigma(1385)^+$. This anisotropy is quantified by fitting a series of Legendre polynomials with the coefficients A$_i$, i= 0, 2 and 4
\begin{equation}
\frac{d\sigma}{d \cos \theta}=A_0 \cdot L_0 + A_2\cdot L_2 + A_4\cdot L_4,
\end{equation}
Only even polynomials are used due to the symmetry in the entrance channel. The results of the fit are listed in the legend of Fig.~\ref{sigmaCM}. The non-zero contribution of the p-wave polynomials $L_2$ already reflects the peripheral character of the production mechanism.
This differential distribution was obtained with acceptance corrections determined by
simulations in which particles are emitted according to phase space. This evaluation of the acceptance will not be sufficient if other, independent angular distributions show a deviation from an isotropic phase-space emission.

The angular distributions of the neutron and $K^+$ were
determined in the same way as for the $\S(1385)$. However, the interdependence of the three precludes further information about
deviations from emission according to phase space.

\paragraph{Angular distributions in the helicity angle frame.}

\begin{figure}[!htb]
\begin{center}
\includegraphics[viewport= 100 20 400 400,angle=0,scale=0.45]{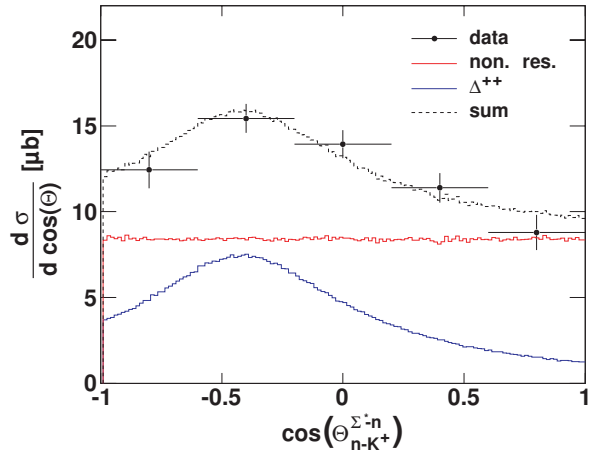}
\caption[]{Color online. Angular differential cross section for the $\Sigma(1385)^+$ production as a function of $cos(\Theta^{\Sigma^*-n}_{nK^+})$. The red and blue histograms correspond to the contributions by the non-resonant production of the  $\Sigma(1385)^+$ and resonant production via an intermediate $\Delta^{++}$, respectively.}
\label{cosTnK_S}
\end{center}
\end{figure}

 We consider now the helicity distributions with respect to $\Theta^{\Sigma^*-n}_{n-K^+}$, i.e. the angle between the $\Sigma(1385)^+$ and the neutron in the neutron-$K^+$ reference frame. For the acceptance corrections, phase space simulations filtered by the CMS angular distribution of the $\Sigma(1385)^+$ (see Fig.~\ref{sigmaCM}) have been used. Since there is no obvious kinematical correlation between the two angles $\Theta^{\Sigma^*-n}_{n-K^+}$ and $\Theta^{\Sigma^*}_{CMS}$, the filtering is not expected to bias the extracted distribution in the helicity frame, but accounts only for the correct acceptance of each event. As can be seen in Fig.~\ref{cosTnK_S}, the helicity angular distribution is not isotropic but shows a peak around $cos(\Theta^{\Sigma*-n}_{n-K^+})=-0.5$. 
This effect must be taken into account to get the appropriate acceptance corrections.\\ 
Higher resonances, contributing to the production of the $\Sigma(1385)^+$, could induce such an effect. In ref.~\cite{Chi68}, the production of $\Sigma(1385)^+$ for the same reaction system as analyzed in the present work, was investigated at a beam momentum of $p=\,6\,\mathrm{GeV/}c$. They found that a part of the $\Sigma(1385)^+$ production proceeds via an intermediate $\Delta^{++}$ resonance with a Breit-Wigner mass peak value of around $2035\, \mathrm{MeV/c}^2$, followed by the decay $\Delta^{++}\rightarrow\Sigma(1385)^++K^+$.
The statistics was not sufficient to obtain any information about the quantum numbers of this $\Delta^{++}$ state, however, the width of the resonance was estimated to be about $250\,\mathrm{MeV}$. These parameters were used as an input for a Monte-Carlo simulation of the process $p+p\rightarrow\Delta^{++}+n$ with subsequent $\Delta^{++}$ decay into $\Sigma(1385) + K^+$. From the simulation  we obtain the shape of the angular distribution of the $\Sigma(1385)$ in the helicity angle frame. Figure~\ref{cosTnK_S} shows, together with the experimental data, simulation results stemming from non-resonant production of the $\Sigma(1385)^+$ (red histogram) and via a $\Delta^{++}$ production (blue histogram). The dashed curve shows the sum of the two simulated processes assuming a contribution of $66\, \%$ by the non-resonant production and $33\, \%$ by the $\Delta^{++}$ excitation. One can see that the agreement with the data is excellent. This result suggests that a rather large amount of the extracted $\Sigma(1385)^+$ may be produced via an intermediate $\Delta^{++}$ resonance. In the following it is assumed that indeed $33\, \%$ of the reconstructed $\Sigma(1385)^{+}$ stems from an intermediate $\Delta^{++}$. 
 \begin{figure*}[!htb]
\begin{center}
\includegraphics[viewport= 30 0 400 460,angle=0,scale=0.96]{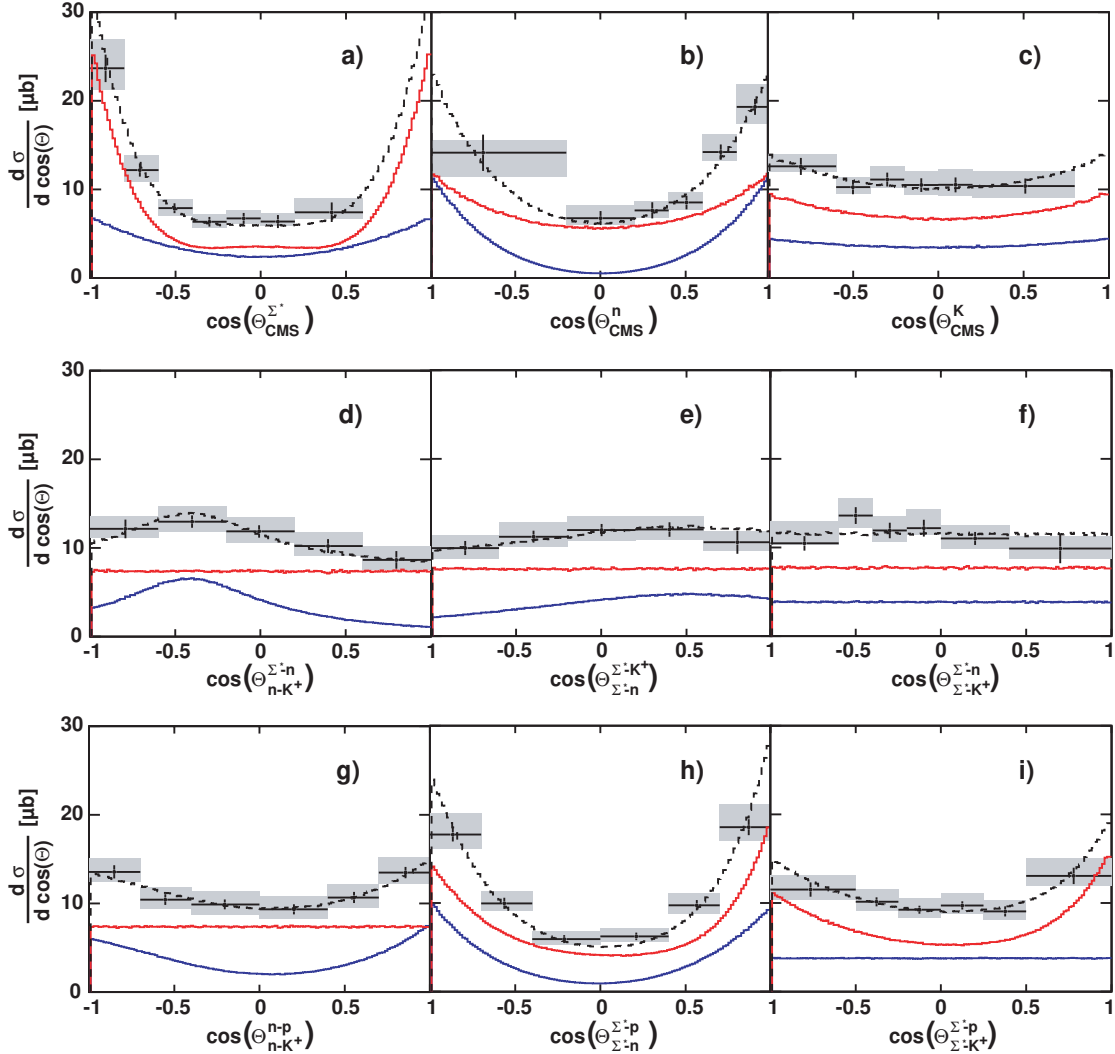}
\caption[]{Angular differential cross sections for the $\Sigma(1385)^+$ production in CMS (top row: $a:\,\Theta^{\Sigma^*}_{CMS},\,b:\,\Theta^{n}_{CMS},\,c:\,\Theta^{K^+}_{CMS}$), helicity (middle row $d:\,\Theta^{\Sigma^*-n}_{n-K^+},\,e:\,\Theta^{\Sigma*-K^+}_{\Sigma^*-n},\,f:\,\Theta^{\Sigma^*-n}_{\Sigma^*-K^+}$) and Gottfried-Jackson angles (bottom row: $g:\,\Theta^{n-p}_{n-K^+},\, h:\,\Theta^{\Sigma^*-p}_{\Sigma^*-n},\,i:\,\Theta^{\Sigma^*-p}_{\Sigma^*-K^+}$) angle frames.
The solid curves represent the contributions by the non-resonant (blue) and resonant (red) production mechanisms, as in Fig.~\ref{cosTnK_S}. The dashed curve shows the sum of the resonant and non-resonant contributions.}
\label{3dist}
\end{center}
\end{figure*}
To obtain the appropriate filter functions for the simulations an iterative procedure is applied. In a first step the differential distribution of the $\Sigma(1385)^+$ in the CMS is used to filter the non-resonant simulations. The differential cross section of the neutron in the CMS is used to filter the $\Delta^{++}$ simulations. This is a natural choice, as the $\Delta^{++}$ and the neutron are going back to back in this reference frame. 
The acceptance corrections are recalculated using the filtered simulation and new experimental distributions are obtained. The process is repeated again and the new filter function is extracted from the experimental distributions. 

The final results after three steps of iteration are shown in Fig.~\ref{3dist}, panel (a) for the $\Sigma(1385)^+$ distribution in the CMS and in panel (b) for the neutron distribution in the CMS. In addition to the experimental data the two components of
the simulated yields are drawn. The red curves represent the non-resonant contribution and the blue curve the $\Delta^{++}$ contribution. 
 The experimental data are shown together with the statistical and systematic errors (gray rectangles). The systematic errors have been evaluated by varying the analysis cuts in the following way. The cuts on the neutron missing mass and on the $\Lambda$ invariant mass have been expanded by $20\,\%$, the cut on the  $K^+$ mass has been varied by $+20\,\%$ and $-10\,\%$. The energy loss cuts in the MDC and TOFino detector have been also varied by $\pm 10\,\%$.
 
 The agreement of the simulated data with the experimental points in panels (a) and (b) is a constraint imposed by the iterative procedure described above, on the other hand the validity of the new acceptance correction is supported by the fact that the distribution in panel (a) is more symmetric then the one shown in Fig.~\ref{sigmaCM}. In order to check whether the simulated model really fits to the experimental data, other, independent angular distributions have to be analyzed. Panel (c) of Fig.~\ref{3dist} shows the CMS angular distribution for the $K^+$ which is also well reproduced by the simulated model.
 Panels (d), (e) and (f) depict the distributions of the three helicity angles and panels (g), (h) and (i) the distributions of the Gottfried-Jackson angles.
 Panel d) shows the distribution of $cos(\Theta^{\Sigma^*-n}_{n-K^+})$ already discussed in Fig.~\ref{cosTnK_S}. The small differences with respect to the previous figure arise by the modified acceptance correction that now takes into account the CMS angular distributions. Panel (e) represents the $cos(\Theta^{\Sigma^*-K^+}_{\Sigma^*-n})$ distribution. There the anisotropy of the experimental distribution due the $\Delta^{++}$ contribution is less evident but still present. The helicity angular distribution (panel (f)) is rather flat for both the experimental and the simulated curves. A possible anisotropy in the experimental data could be connected to the polarization of the $\Delta^{++}$ that was not included in the simulations. The data leaves some room for anisotropy but no firm conclusion can be drawn.
 \paragraph{Angular distributions in the Gottfried-Jackson frame.}
 The experimental distributions of the Gottfried-Jackson angles show a strong anisotropy that can be well reproduced by our simple model. The distributions shown in panels (g) and (i) of Fig.~\ref{3dist} are linked to the partial waves acting at the respective verteces but no specific assumption is made in our simulation. Nevertheless, these distributions are kinematically correlated with the CMS distributions. This is also at the origin of the good agreement between the experimental and the simulated angular distributions in the helicity frame.    
 
 It is clear that the nine angular distributions are not all kinematically independent from each other, they are here shown for the sake of completeness. These distributions constitute an important reference to test future theoretical models describing the production. 
 Indeed different production mechanisms will result in different angular distributions of the decay products.
 It has also to be pointed out that the internal degrees of freedom of the $\Sigma(1385)^+$ and the intermediate $\Delta^{++}$ resonance are neglected in our simulations.
 
 The overall agreement of the experimental data with the angular distributions modeled with the simulations justifies the usage of the resulting acceptance corrections, despite of the fact that the contribution by the $\Delta^{++}$ resonance cannot be demonstrated unambiguously. 
 \subsection{Production cross sections}
The last step of the analysis consists of the calculation of the production cross section. The production cross section has been estimated  by the integration of the simulated differential cross sections adapted to the experimental data over the whole phase space for each of the nine distributions shown in Fig.~\ref{3dist}. Indeed, assuming that proper acceptance and efficiency corrections have been applied, the same total cross section for the exclusive production of the $\Sigma(1385)^+$ resonance in the reaction pp $\rightarrow n+ K^+ +\Sigma^+(1385)$ should be obtained.   

\begin{figure}[!htb]
\begin{center}
\includegraphics[viewport= 50 0 500 400,angle=0,scale=0.45]{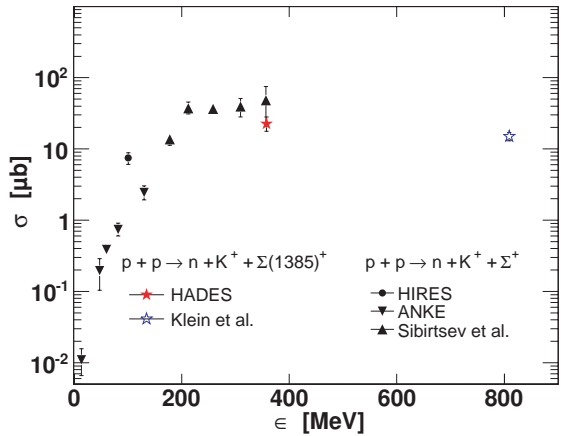}
\caption[]{Color online. Production cross sections for the reaction $p+p\rightarrow p+K^++\Sigma^+$ as a function of the excess energy $\epsilon$. Data points measured by several experiments have been compiled in \cite{Bud10}. The production cross section for the reaction $p+p\rightarrow n+K^++\Sigma(1385)^+$ from this work (red star) and at $6\,\mathrm{GeV/c}$ incident momentum \cite{Kle70} (blue star) are shown as well. }
\label{cross}
\end{center}
\end{figure}

The integration of the nine differential angular distributions shown in Fig.~\ref{3dist} delivers values of the total production cross section that are in agreement within the statistical errors. The arithmetic mean of these values yields $22.42 \pm0.99 \pm 1.57  ^{+3.04}_{-2.23}~\mu \mathrm{b}$. The statistical error is followed by a first systematic error arising from the normalization to the elastic events and a second asymmetric error stemming from the cut variations discussed above.
This cross section can be compared to the values known for the $\Sigma(1192)^+$ production in p+p collisions.
Figure~\ref{cross} shows the production cross sections for the reaction $p+p\rightarrow \Sigma^+ + K^+ +n$ as a function of the excess energy $\epsilon$. The cross section extracted from our analysis of the channel $p+p\rightarrow \Sigma(1385)^+ + K^+ +n$  is shown together with the measurement at higher energies reported in \cite{Kle70}.

The data point corresponding to the $\Sigma(1385)^+$ is about a factor two lower than the cross section value extracted for the $\Sigma^{+}$  at the same excess energy but still compatible within the systematic errors.
Our data point agrees well with the value of $15\pm2\,\mu \mathrm{b}$ reported in \cite{Kle70} for $\Sigma(1385)^+$ at an excess energy of $830\,\mathrm{MeV}$ also in p+p collisions, pointing to a weak dependence on excess energy of both $\Sigma^+$ and $\Sigma(1385)^+$ production above $\epsilon=\,200\mathrm{MeV}$.

\section{Summary}
We present the results obtained from an exclusive analysis of the $\Sigma(1385)^+$ resonance produced in p+p collisions at a kinetic energy of 3.5 GeV. The sophisticated method developed for the reconstruction of the background allows to estimate with high precision the position and the width of this resonance. The value extracted for the mass, $m_0=1383.2\pm 0.9 ^{+0.1}_{-1.5}\, \mathrm{MeV/}c^2$, agrees well with the PDG value \cite{pdg}, while the width, $\Gamma_0=40.2\pm2.1^{+1.2}_{-2.8}\,\mathrm{MeV/}c^2$, is about $4\, \mathrm{MeV/}c^2$ larger than the PDG average \cite{pdg}.

Angular distributions have been corrected for acceptance and detector response by means of a complete simulations testing two assumptions for the production mechanism.
Our analysis suggests that a $33\,\%$ of the $\Sigma(1385)^+$ yield originates from the decay of an intermediate $\Delta^{++}$ resonance, whereas one should underline that interference effects could also modify the kinematics of the reaction. The efficiency and acceptance corrected angular distribution of the three reference frames CM, helicity and Gottfried-Jackson have been extracted and can be used to test theoretical models describing the production mechanisms for the $\Sigma(1385)^+$ state. A total production cross section of $\sigma=\, 22.42 \pm0.99 \pm 1.57  ^{+3.04}_{-2.23}~\mu \mathrm{b}$ has been deduced and is found to be consistent with the systematics measured for the ground state $\Sigma^+$ production in the same final state as a function of the excess energy. The available data base do not exhibit a strong dependence of the production cross section on the excess energy in the range $200\leq\,\epsilon\leq\,800\,\mathrm{MeV}$ for both the ground state $\Sigma^+$ and the first excited state $\Sigma(1385)^+$.
Our results provide the necessary reference for further studies of the spectral shape of the $\Sigma(1385)$ resonance in p+A and A+A reactions. Furthermore, our analysis represents an important bench mark for the investigation of the $\Lambda(1405)$ resonance in p+p, p+A and A+A collisions.
Our results will help to understand the properties and production characteristics of the lowest lying hyperon resonances both 
in elementary and heavy ion collisions with proton induced reactions at nuclei as link between them.  
\subsection*{Acknowledgements}
The authors would like to thank Dr. B. Ketzer and Prof. W. Weise for the useful discussions.
The following funding are acknowledged.
LIP Coimbra,
Coimbra (Portugal): PTDC/FIS/113339/2009,
SIP JUC Cracow, Cracow (Poland): NN202286038, NN202198639, HZ Dresden-Rossendorf, Dresden
(Germany): BMBF 06DR9059D, TU Muenchen,
Garching (Germany) MLL Muenchen DFG EClust:
153 VH-NG-330, BMBF 06MT9156 TP5 TP6, GSI
TMKrue 1012, GSI TMFABI 1012, NPI AS CR,
Rez (Czech Republic): MSMT LC07050, GAASCR
IAA100480803, USC - S. de Compostela, Santiago de
Compostela (Spain): CPAN:CSD2007-00042, Goethe
Univ. Frankfurt (Germany): HA216/EMMI, HIC
for FAIR (LOEWE), BMBF06FY9100I, GSI F\&E01,
CNRS/IN2P3 (France).

\end{document}